\documentclass[twocolumn,showpacs,preprintnumbers,amsmath,amssymb]{revtex4}

\usepackage{graphicx}
\usepackage{dcolumn}
\usepackage{bm}
\usepackage{mathrsfs}
\usepackage{color}

\newcommand{\braket}[1]{ \langle #1 \rangle}
\newcommand{\bra}{ \langle }
\newcommand{\ket}{\rangle}
\newcommand{\nonum}{\nonumber \\}
\newcommand{\bv}[1]{\mbox{\boldmath $#1$}}

\begin{document}
\title{ Precise comparison of the Gaussian expansion method and the Gamow shell model 
}

\author{H. Masui}
\email{hgmasui@mail.kitami-it.ac.jp}
\affiliation{%
 Information Processing Center,
  Kitami Institute of Technology, Kitami
 090-8507, Japan
}%

\author{K. Kat\=o}
\affiliation{
  Nuclear Reaction Data Centre, Faculty of Science,
  Hokkaido University, Sapporo 060-0810, Japan
}%

\author{N. Michel}
\affiliation{
National Superconducting Cyclotron Laboratory,
Department of Physics and Astronomy, Michigan State University,
East Lansing, MI 48824 USA \\
Grand Acc\'el\'erateur National d'Ions Lourds (GANIL), CEA/DSM -- CNRS/IN2P3, BP 55027,  F-14076 Caen Cedex, France  
}%

\author{M. P{\l}oszajczak}
\affiliation{
  Grand Acc\'el\'erateur National d'Ions Lourds (GANIL), CEA/DSM -- CNRS/IN2P3, BP 55027,  F-14076 Caen Cedex, France  
}%

\date{\today}

\begin{abstract}

We perform a detailed comparison of results of the Gamow Shell Model (GSM)
and the Gaussian Expansion Method (GEM) supplemented
by the complex scaling (CS) method for the same
translationally-invariant cluster-orbital shell model (COSM) Hamiltonian.
As a benchmark test, we calculate the ground state $0^{+}$ and the first excited state $2^{+}$ of mirror nuclei $^{6}$He and $^{6}$Be in the model space consisting of two valence nucleons in $p$-shell outside of a $^{4}$He core. We find a good overall agreement of results obtained in these two different approaches, also for many-body resonances.
\end{abstract}


\maketitle

\section{Introduction}

In recent years, the playground of nuclear physics has extended
towards neutron and proton drip lines~\cite{Ta85_PRL85,Oz01_NPA691,Oz02_NPA693}.
Huge amount of new experimental data on nuclei far from the valley of
stability has been provided by new rare-isotope facilities.
The knowledge of these nuclei has largely improved also
due to the progress in theoretical methods and computing power
which allows to calculate light nuclei in {\em ab initio} framework
taking into account the proximity of the scattering continuum.
The description of various manifestations of the continuum coupling
requires the generalization of existing many-body methods
and call for theories which unify structure and reactions.

Realistic studies of the coupling to continuum in the many-body
framework can be made in the open quantum system extension of the
Shell Model (SM), the so-called Continuum Shell Model
(CSM)~\cite{[Oko03],CSM_Volya}. A recent realization of the CSM is the
complex-energy CSM based on the Berggren ensemble~\cite{Ber68}, the
GSM, which finds a mathematical setting in the Rigged Hilbert
Space~\cite{Gel61}. This model is a natural generalization of the
standard  SM for the description of configuration mixing in weakly
bound states and resonances.  Berggren completeness relation can be
derived from the Newton completeness  relation~\cite{Ne60_JMP1}
for the set of real-energy eigenstates by deforming the real momentum
axis to include resonant poles which are located in the fourth
quadrant of the complex $k$-plane. Thus the Berggren completeness
relation which replaces the real-energy scattering states by the
resonance contribution and a background of complex-energy continuum
states, puts the resonance part of the spectrum on the same footing as
the bound and scattering spectrum. As the benefit of the explicit
inclusion of the non-resonant continuum and resonant poles, the
contribution of the unbound states to the one- and two-body matrix
elements can be discussed. Berggren ensemble has found the application
in the GSM~\cite{PRL89},
time-dependent Green's function approach~\cite{Vo09_PRC79},
the no-core GSM~\cite{papa},
the coupled cluster approach~\cite{CC07},
the Density Matrix Renormalization Group (DMRG) approach~\cite{Rot06},
and in the coupled-channel GSM~\cite{Jag12,Be14_PLB730} to study 
various nuclear structure and reaction problems. 

Another approach is the complex scaling (CS) method~\cite{Ag71_CMP22},
which has been used to solve many-body resonances
in many fields including atomic physics, molecular physics~\cite{Ho83,Mo98}
and nuclear physics~\cite{Ga86_PRC34,Ao06}.
In the CS method, asymptotically-divergent 
resonant states are described within ${\cal L}^{2}$-integrable functions
through the rotation of space coordinates and their conjugate momenta
in the complex plane. As basis functions,
the Gaussian Expansion Method (GEM)~\cite{Hi03_PPNP21}
has been extensively employed for
the cluster-orbital shell model (COSM)~\cite{Su88_PRC38}
and coupled rearrangement channel model such as the TV-model~\cite{Ao95_PTP93}.
The CS-COSM has successfully been applied to description of
resonant states observed above the many-body decay threshold
in $p$-shell nuclei ($A=5$-$8$) using a $^4$He+$XN$ model,
where $X=1$-$4$ and $N=p$, $n$~\cite{My10,My11}.
The CS-TV model for the core+2N systems has been shown
to reproduce the observed Coulomb breakup cross sections
for three-body continuum energy states~\cite{Ki10,Ki13}.    

The purpose of these studies is to perform a detailed comparison of
the GSM and the GEM+CS results for $^6$He and $^6$Be using the same
COSM coordinates for valence nucleons~\cite{Su88_PRC38} and the same 
Hamiltonian. In COSM, all coordinates are taken with respect to the core Center-of-Mass (CoM),
so that the translational invariance is strictly preserved. COSM
combined with CS method has been employed in numerous studies of
weakly bound states and resonances in light
nuclei~\cite{Ao95_PTP93,My98_PTP99,My01_PRC63,My02_PTP108,Ao06_PTP116,Ma06_PRC73,Ma07_PRC75,Ma09_EPJA42,Ma12_NPA895}.
COSM coordinates have been also used in GSM~\cite{vertse} to investigate
isospin mixing in mirror nuclei~\cite{iso} and charge radii in halo nuclei~\cite{radii}.

The paper is organized as follows.
In Section II we present our COSM Hamiltonian and the model space.
In Section III, the two theoretical approaches,
namely the GEM+CS (Section III.A) and the GSM (Section III.B),
are briefly introduced. GEM+CS and GSM results for $^6$He  and $^6$Be
are presented and discussed in Section IV.
Finally, Section V gives the main conclusions of these studies.

\section{The COSM Hamiltonian}

\begin{figure}[htpb]
  \includegraphics{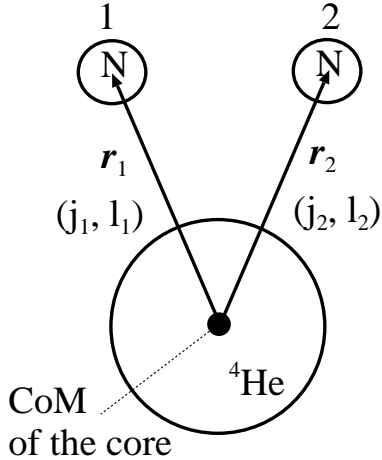}
    \caption{Coordinate system of the COSM approach.}
    \label{fig:COSM_co}
\end{figure}

In these studies, we employ the three-body model for $^{4}$He plus
two-nucleon system in the COSM coordinates~\cite{Su88_PRC38} (see Fig.~\ref{fig:COSM_co}).

The Hamiltonian is written as follows:
\begin{equation}
  \label{eq:Hamiltonian_01}
  \hat{H} = \sum_{i=1}^{2}
  \left(
    \hat{t}_{i} + \hat{V}_{i}^{(C)}
  \right)
  +\left(
    \hat{T}_{12}
  + \hat{v}_{12}
  + \hat{V}_{12}^{(C)}
  \right)
\mbox{ ,}  
\end{equation}
where $\hat{t}_{i}$ and $\hat{V}_{i}^{(C)}$ are the kinetic
and potential energy operators for the $^{4}$He core
and an $i$th valence nucleon subsystem.
In Eq.~(\ref{eq:Hamiltonian_01}), the first parenthesis corresponds to the 
single-particle Hamiltonian for the $i$th valence nucleon, which is defined as
\begin{eqnarray}
  \label{eq:Hamiltonian_03}
  \hat{h}_{i} & \equiv & \hat{t}_{i} + \hat{V}_{i}^{(C)}
  \mbox{,}
  \hspace{3mm}
  (i = 1, 2)
  \mbox{ .}
\end{eqnarray}
In the second parenthesis of Eq.~(\ref{eq:Hamiltonian_01}),
$\hat{v}_{12}$ is the nucleon-nucleon interaction for valence particles, and:
\begin{equation}
  \label{eq:Hamiltonian_02}
\hat{T}_{12} = -\frac{\hbar^{2}}{M^{(C)}} \nabla_{1} \cdot \nabla_{2}
\mbox{ ,}
\end{equation}
is the recoil part which comes from the subtraction of the center of
mass (CoM) motion, due to the finite mass $M^{(C)}$ of the core nucleus.
The last term of Eq.~(\ref{eq:Hamiltonian_01}) is the three-body potential of
$^{4}$He and two valence nucleons.

The interaction $\hat{V}_{i}^{(C)}$ between the core
and the $i$th valence nucleon contains three terms: 
\begin{equation}
  \label{eq:pot_core_N}
  \hat{V}_{i}^{(C)}  = \hat{V}^{\alpha n}_{i}
  + \hat{V}_{i}^{\rm Coul}
  + \lambda \, \hat{\Lambda}_{i}
  \mbox{ , }
  \hspace{3mm}
  (i = 1, 2 )
  \mbox{ .}
\end{equation}
The nuclear interaction part $\hat{V}^{\alpha n}_{i}$ 
is the modified KKNN  potential~\cite{Ka78_PTP61,Ao95_PTP93},
which reproduces the $\alpha$-N phase shifts in the low energy region.
This potential contains a central and an $LS$ parts as
\begin{eqnarray}
  \label{eq:KKNN_all}
  \hat{ V}^{\alpha n}_{i} (r_{i}) & = & 
    V^{\alpha n}_{0}(r_{i})
  +  2 V^{\alpha n}_{LS}(r_{i}) \, \bv{L}\cdot\bv{S}
  \mbox{,}
  \hspace{3mm}
  ( i = 1, 2 )
  \mbox{ ,}
\end{eqnarray}
where $\bv{r}_{i}$ is the relative coordinate between $^{4}$He and the
$i$th valence nucleon. The central part of Eq.~(\ref{eq:KKNN_all}) is written as:
\begin{eqnarray}
  \label{eq:KKNN_central}
   V^{\alpha n}_{0}(r_{i})
   & = &  \sum_{k=1}^{5}\, [(-1)^{\ell_{i}}]_{k} \,
   V_{k}^{0} \, \exp(-\rho^{0}_{k} \,r^{2}_{i})
\mbox{ ,}  
\end{eqnarray}
where $[(-1)^{\ell_{i}}]_{k}$ is given by:
\begin{equation}
  \label{eq:k_dep}
   [(-1)^{\ell_{i}}]_{k} = 
  \left\{
  \begin{array}{cl}
    1 & ( k=1, 2) \\
    (-1)^{\ell_{i}} & (k=3, 4, 5)
  \end{array}
  \right.
  \mbox{ .}
\end{equation}
The $LS$ part is:
\begin{eqnarray}
  \label{eq:KKNN_LS}
  V^{\alpha n}_{LS}(r_{i})
 & = &
 \sum_{m=1}^{3} \, f^{LS}_{m} \,
  V_{m}^{LS} \, \exp(-\rho_{m}^{LS} \,r^{2}_{i}) 
  \mbox{ ,}
\end{eqnarray}
where the factor $f^{LS}_{m}$ is:
\begin{equation}
  \label{eq:k_dep_ls}
   f^{LS}_{m} = 
  \left\{
  \begin{array}{cl}
    1 & ( m=1) \\
    1 - 0.3\times(-1)^{\ell_{i}} & (m=2, 3)
  \end{array}
  \right.
  \mbox{ .}
\end{equation}
Parameters of the modified KKNN potential~\cite{Ka78_PTP61,Ao95_PTP93}
are shown in Table~\ref{tab:KKNN_param}.
\begin{table}
  \centering
  \caption{Parameters of the modified KKNN
    potential~\cite{Ka78_PTP61,Ao95_PTP93}
    used in this calculation.}
  \begin{tabular}{rrrrrrrr}
   \hline
    \hline
    $k=$ & 1 & 2 & 3 & 4 & 5 \\
    \hline
    $ V_{k}^{0} $ [MeV]  &
    $-96.3$ &  $77.0$  &  $34.0$ &  $-85.0$ &  $51.0$ \\
    $\rho_{k}^{0}$ [fm$^{-2}$] &
    $0.36$ &  $0.90$ & $0.20$ & $0.53$ & $2.50$ \\  
    \hline
   $ V_{k}^{LS} $ [MeV] & $-8.4$ & $-10.0$ &  $10.0$ & --- & ---\\
    $\rho_{k}^{LS}$ [fm$^{-2}$] & $0.52$  & $0.396$ & $2.20$ & --- & ---\\
    \hline
    \hline
  \end{tabular}
  \label{tab:KKNN_param}
\end{table}

For the Coulomb part $\hat{V}_{i}^{\rm Coul}$ in
Eq.~(\ref{eq:pot_core_N}),
we use a folded-type Coulomb interaction for the $^{4}$He+$p$ subsystem:
\begin{equation}
  \label{eq:Coulomb_Core_p}
  \hat{V}^{\rm Coul}_{i}(r_{i}) = \frac{2e^{2}}{r_{i}} \,
  \mbox{Erf}(\alpha \, r_{i})
    \mbox{ ,}
\end{equation}
where $ \mbox{Erf}(r)$ is the error function, and $\alpha = 0.828$ fm$^{-1}$.

To eliminate the spurious states in the relative motion between
$^{4}$He-core and the valence nucleon in CS,
we use a projection operator~\cite{Sa77_PTPS62}:
\begin{equation}
  \label{eq:OCM_01}
  \hat{\Lambda}_{i} = \lambda | FS \ket \bra FS |
  \mbox{ ,}
\end{equation}
where the forbidden state in the $^{4}$He+$N$ case; $| FS \ket = | 0s_{1/2} \ket$,
is given by the harmonic oscillator function with the size parameter $b=1.4$ fm. 

In GSM, the forbidden state is eliminated from the set of
the single-particle states, $\phi_{i}$, as 
\begin{equation}
  \label{eq:GSM_OCM_01}
  \phi_{i} \Rightarrow (1 - \hat{\Lambda}_{i}) \phi_{i}
  \mbox{ .}
\end{equation}

We can confirm that the core-particle potential (\ref{eq:pot_core_N})
with the parameters given in Table~\ref{tab:KKNN_param},
reproduces experimental energies and widths of $3/2_{1}^{-}$
and $1/2_{1}^{-}$ resonances in
the $^{5}$He($^{4}$He+$n$) and $^{5}$Li( $^{4}$He+$p$) systems.

For the two-body interaction $\hat{v_{12}}(\bv{r}_{12})$
of valence nucleons, 
where $\bv{r}_{12} \equiv \bv{r}_{1} - \bv{r}_{2}$,
we use the Minnesota potential~\cite{Th77_NPA286}:
\begin{eqnarray}
  \label{eq:Minnesota_01}
  & & \hat{v}_{12}(\bv{r}_{12})   \nonum
  & & =  
  \sum_{k=1}^{3} \, V_{k}^{0} \,
  \left(
    W^{(u)}_{k} - M^{(u)}_{k} P^{\sigma} P^{\tau} 
   +B^{(u)}_{k} P^{\sigma}  - H^{(u)}_{k} P^{\tau} 
 \right)
  \nonum
  & &
  \hspace{8mm} \times
  \exp(-\rho_{k} \,\bv{r}^{2}_{12})
 \mbox{ .}
\end{eqnarray}
Parameters of this interaction are summarized in Table \ref{tab:Minnesota_param}, 
and the exchange parameter is taken as $u=1.0$.
\begin{table}
  \centering
  \caption{Parameters of the Minnesota potential~\cite{Th77_NPA286}.}
  \label{tab:Minnesota_param}
  \begin{tabular}{ccccccccccc}
    \hline
    \hline
    $k$ && 1 && 2 && 3 \\
    \hline
     $V_{k}^{nn}$[MeV]  &&
    $  200$ &&  $ -178 $ & &  $-91.85 $  \\
    $\rho_{k}^{nn}$[fm$^{-2}$] &&
    $ 1.487 $ &&  $0.639  $ && $  0.465$\\
    \hline
    $W^{(u)}_{k}$ && $u/2$ && $u/4$ && $u/4$   \\
    $M^{(u)}_{k}$ && $(2-u)/2$ && $(2-u)/4$& & $(2-u)/4$   \\
    $B^{(u)}_{k}$ && $0$ && $u/4$ & & $-u/4$   \\
    $H^{(u)}_{k}$ && $0$ && $(2-u)/4$ && $-(2-u)/4$   \\
    \hline
    \hline
  \end{tabular}
\end{table}
The Coulomb interaction between valence protons is taken as an ordinary $1/r$-type 
functional form.

It was shown that the binding energy of $^{6}$He cannot be reproduced
using the reliable one- and two-body potentials
for core-particle and particle-particle parts,
respectively~\cite{Ao95_PTP93,My01_PRC63}.
The correct binding energy in a system $^{4}$He+$N$+$N$
is recovered by using a simple two-body Gaussian interaction, mimicking a physical 
three-body effect in the system~\cite{My01_PRC63} as: 
\begin{equation}
  \label{eq:V-cnn}
  \hat{V}_{12}^{\rm (C)}(r_{1}, r_{2}) = V^{0}_{\alpha nn} \,
  \exp(-\rho_{\alpha nn} (r^{2}_{1}+r^{2}_{2})) 
\end{equation}
with the parameters $V^{0}_{\alpha nn} = -0.41$ MeV
and $\rho_{\alpha nn} = 5.102 \times 10^{-3}$ fm$^{-2}$.

\section{The models}

In this section, we discuss two models for solving 
$^{4}$He+$2N$ ($N$ is proton or neutron) systems with the COSM Hamiltonian.
One is the GEM+CS approach, and another one is the GSM approach.
The essential differences between the GEM+CS and GSM approaches are
the choice of the basis functions and the treatment of continuum states.

The basis function  $\Phi(\bv{r}_{1} , \bv{r}_{2} )$
in COSM is defined with a product of the functions
with respect to each coordinate from the core to a valence nucleon,
\begin{eqnarray}
  \label{eq:Basis_001}
 \Phi(\bv{r}_{1} , \bv{r}_{2} )_{JM} 
  &  \equiv &
  [{\cal A}
  \left\{
  \phi_{\alpha_{1} }(\bv{r}_{1}) 
    \otimes
    \phi_{\alpha_{2} }(\bv{r}_{2})
  \right\}
    ]_{JM}
  \mbox{ .}
  \nonum
\end{eqnarray}
Here, $\alpha_{i}$ denotes the angular part of the $i$th
particle $\{ j_{i}, \, \ell_{i}  \}$, and its $z$-components are implicitly included.
${\cal A}$ is the antisymmetrizer for particles 1 and 2.

The basis function for the $i$th valence nucleon is 
\begin{equation}
  \label{eq:Basis_002-1}
  \phi_{\alpha_{i} }(\bv{r}_{i})
  = f(r_{i}) | j_{i} m_{i} \ket
  \mbox{ .}
\end{equation}
The angular momentum part of the basis function is constructed
by using the normal $jj$-coupling scheme as
\begin{eqnarray}
  \label{eq:Basis_002}
  | J M \ket & = &
  | [j_{1} \otimes j_{2}]_{JM} \ket
  \mbox{ .}
\end{eqnarray}
The above coupling procedure is the same both for GEM and GSM.

\subsection{The Gaussian expansion method with complex scaling}

The radial part of the GEM wave function is not an eigenfunction of the single-particle
Hamiltonian $\hat{h}_{i}$, but the Gaussian function with the width parameter $a$ as
\begin{eqnarray}
  \label{eq:Basis_004}
  f_{n_{i}}(r_{i}) & \equiv & u^{n_{i}}_{\ell_{1}}(r_{i})
  \nonum
   & = &  N_{i} r_{i}^{ \, \ell_{i}}
  \exp(-\frac{1}{2} a_{n_{i}} r_{i}^{2})
  \mbox{ ,}
  \hspace{3mm}
  (i = 1, 2)
  \mbox{ ,}
\end{eqnarray}
where $N_{i}$ is the normalization, and $\ell_{i}$ is angular momentum for the $i$th nucleon.

The width parameter $a_{n_{i}}=1/b_{n_{i}}^{2}$
in the GEM basis functions is defined using the geometric progression as:
$b_{n_{i}} = b_{0} \gamma^{n_{i}-1}$~\cite{Hi03_PPNP21}.
Here, $b_{0}$ and $\gamma$ are input parameters,
and $n_{i}$ is an integer. The model space of the system is spanned by basis functions 
from $n_{i}=1$ to $N_{\rm max}$. The $k$th eigenfunction $ \psi_{k:\, \alpha_{i} }$
of the core+$N$ system can be obtained by diagonalizing the
single-particle Hamiltonian $\hat{h}_{i}$ with the Gaussian basis functions,
\begin{equation}
  \label{eq:Basis_006}
  \psi_{k:\, \alpha_{i}} (\bv{r}_{i})
= \sum_{m}^{N_{\rm max}} \, c_{m}^{(k)} \phi^{(m)}_{\alpha_{i}} (\bv{r}_{i})
\mbox{ .}
\end{equation}
Here, $\hat{h}_{i} \psi_{k:\, \alpha_{i} } = \epsilon_{k}
\psi_{\alpha_{i}}$, and $c_{m}^{(k)}$ are determined by using the
variational principle.

For solving the core+$2N$ system, the basis function (\ref{eq:Basis_001}) 
is given by the product of basis functions in Eq.~(\ref{eq:Basis_004})
for particle 1 and 2 as follows:
\begin{eqnarray}
  \label{eq:Basis_007}
  \Phi^{(m)}_{JM}
  & = &
  {\cal A}
  \left\{
     u_{\ell_{1}}^{(m)}(r_{1}) \cdot
     u_{\ell_{2}}^{(m)}(r_{2})
 | J M \ket^{(m)}     
\right\}
\mbox{ .}
\nonum
\end{eqnarray}
Here, the width parameters $a_{i}^{(m)}$ in $ u_{\ell_{i}}^{(m)}$ 
are prepared independently for particle 1 and 2.
$m$ is the index of the one-body basis functions.

\begin{figure}[htpb]
  \includegraphics{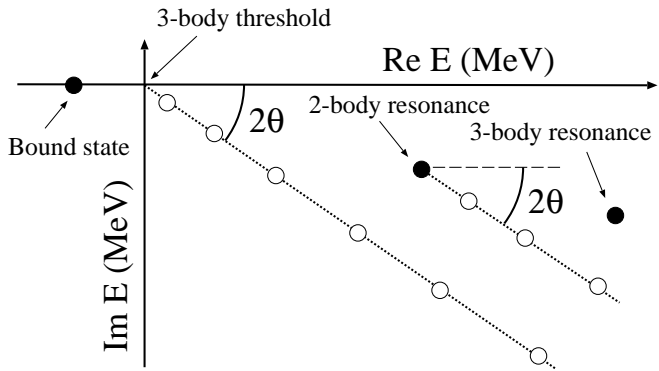}
    \caption{Complex-scaled eigenstates of the three-body
      Hamiltonian for the Borromean system.
      Solid circles are bound and resonance states,
      and open circles are continuum states.}
    \label{fig:CSM_E_fig}
\end{figure}
The calculation of two-body matrix elements (TBME),
$\braket{ \Phi^{(m)} | \hat{O}_{12} | \Phi^{(n)} }$
can be performed analytically. Even for different Gaussian width parameters,
we can obtain the value of TBME without any approximations.

The solution of the core+$2N$ system
can be obtained by diagonalizing the Hamiltonian
\begin{equation}
  \label{eq:Basis_009}
  \hat{H} \Phi_{k: \, JM}  =  E_{k} \, \Phi_{k: \, JM}
  \mbox{ ,}
\end{equation}
and the corresponding eigenfunction
is expressed  as a linear combination of the basis functions,
\begin{eqnarray}
  \label{eq:Basis_008}
  \Phi_{k: \, JM} & = &
  \sum_{m}^{N_{\rm Tot}}\,
  C_{m}^{(k)} \, \Phi^{(m)}_{JM}
  \mbox{ .}
\end{eqnarray}

In order to treat the many-body resonant states, we apply the CS method. 
In this method, the coordinate and momentum are transformed
using a rotation angle $\theta$ as:
\begin{equation}
  \label{eq:Basis_013}
  r \rightarrow r \, e^{ i \theta}
  \hspace{3mm}
  (k \rightarrow k \, e^{ -i \theta})
  \mbox{ .}  
\end{equation}

Resonance wave functions, which diverge in the asymptotic region,
can be converged with this transformation for a suitable rotation angle.
This essential feature is proven by the ABC-Theorem~\cite{Ag71_CMP22,Ba71_CMP22}.
After this transformation, all continuum states are aligned along the rotated axis.
Furthermore, using GEM, the continuum states are automatically discretized
through the diagonalization of the Hamiltonian.
A schematic figure of the bound states and resonances and
discretized continuum states are shown for the Borromean system like
$^{4}$He+$N$+$N$ in Fig.~\ref{fig:CSM_E_fig}.

\subsection{Gamow shell model approach}

\begin{figure}[htpb]
  \includegraphics{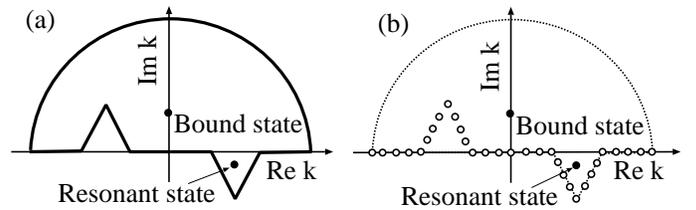}
  \caption{Deformed contour on the complex momentum plane (a),
    and discretized continuum states along the deformed contour (b).
    Solid circles are bound and resonant states, and
    open circles are discretized continuum states.}
    \label{fig:GSM_k_fig}
\end{figure}

Another approach to study many-body resonances
is the GSM approach~\cite{PRL89,papa,CC07,Rot06}.
This generalization of the nuclear SM treats single-particle
bound, resonance and continuum states on the same footing
using a complete Berggren single-particle basis~\cite{Ber68}:
\begin{eqnarray}
  \label{eq:GSM_comp_01}
  \bv{1} & = &  \sum_{i \in b, r} | \phi_{i} \ket \bra \phi_{i} |
  + \oint_{\Gamma_{k}} dk  | \phi(k) \ket \bra \phi(k) | 
  \mbox{ ,} 
\end{eqnarray}
where $\Gamma_{k}$ is a deformed momentum contour.
For each $(\ell,j)$ of the resonant single-particle state in the basis,
the set $(\ell,j)_{\rm c}$ of continuum states along the discretized
contour in $k$-plane enclosing the resonant state(s)
$(\ell,j)$ is included in the basis (see Fig.~\ref{fig:GSM_k_fig}): 
\begin{eqnarray}
  \label{eq:GSM_comp_011}
  \bv{1}
  & \simeq &  \sum_{i \in b, r} | \phi_{i} \ket \bra \phi_{i} |
  + \sum_{\eta \in {\rm cont} } | \phi(k_{\eta}) \ket \bra \phi(k_{\eta}) |
  \mbox{ ,} 
\end{eqnarray}
where $k_{\eta}$ are linear momenta discretized
on the deformed contour with the parameters of
a maximum $k$ and a number of discretized points. 
Different shapes of $(\ell,j)$-contours are equivalent unless
the number of resonant states contained in them changes. 
The complete many-body basis is then formed by all Slater determinants
where nucleons occupy the single-particle states of a complete Berggren ensemble~\cite{PRL89}.

In the Berggren basis, the basis function of the core+$2N$ system is:
\begin{eqnarray}
  \label{eq:Basis_011}
  \Phi^{(\nu)}_{JM}
  & = &
  {\cal A}
  \left\{
    [
     \phi_{1}^{(\nu)} \otimes
     \phi_{2}^{(\nu)}
    ]_{J M}
\right\}
\mbox{ .}
\nonum  
\end{eqnarray}
Here, $\phi_{i}^{(\nu)}$ are single-particle bound, resonance,
and discretized-continuum  states for particles 1 and 2.

In GSM, the deformed contour for each $(\ell,j)$ is varied to obtain
the best numerical precision of calculated eigenenergies and
eigenvalues for a given discretization of the contour.
Since the direct calculation of the TBME using the continuum 
and/or resonant single-particle states is numerically demanding,
and even difficult to define from a theoretical point
of view for some particular instances, one calculates TBME using the harmonic oscillator (HO)
expansion procedure~\cite{vertse}. For the TBME between GSM basis functions 
$ \Phi^{(i)}_{\rm GSM} $ and $ \Phi^{(j)}_{\rm GSM}$, one obtains:
\begin{eqnarray}
  \label{eq:TBME_GSM_01}
  & &  \bra
    \Phi^{(i)}_{\rm GSM} | \hat{O}_{12} | \Phi^{(j)}_{\rm GSM} 
    \ket  \nonum
    & & \nonum
   & = &
   \sum_{\alpha, \beta}
   \bra
   \Phi^{(i)}_{\rm GSM} |
   \Phi_{\rm HO}^{(\alpha)}
   \ket
   \bra
   \Phi_{\rm HO}^{(\alpha)} \, |
   \hat{O}_{12} |
   \Phi_{\rm HO}^{(\beta)}
   \ket
   \bra
   \Phi_{\rm HO}^{(\beta)} |
   \Phi^{(j)}_{\rm GSM} 
   \ket  \nonum
   & = & 
   \sum_{\alpha, \beta}
   d^{*}_{i,\alpha} \,
   d_{j,\beta}
   \bra
   \Phi_{\rm HO}^{(\alpha)} \, |
   \hat{O}_{12} |
   \Phi_{\rm HO}^{(\beta)}
   \ket
    \mbox{ ,}
\end{eqnarray}
where $ \Phi_{\rm HO}^{(\alpha)} $ are HO basis functions and $d_{i,\alpha}$ is the overlap
between the GSM basis function $\Phi^{(i)}_{\rm GSM}$ and the HO basis function:
\begin{equation}
  \label{eq:GSM_comp_02}
d_{i,\alpha} \equiv \bra
   \Phi_{\rm HO}^{(\beta)} \, |
    \Phi^{(i)}_{\rm GSM} 
   \ket
   \mbox{ .}
\end{equation}
The advantage of this procedure is that the TBMEs with the HO
expansion can be stored for a fixed $b_{\rm HO}$,
and one only needs to calculate the overlaps $d_{i,\alpha}$,
whatever the Berggren states are.

\section{Results}

For numerical calculations, we define the number of basis states.
In the GEM+CS approach, the number of radial wave functions 
for each valence nucleon $N_{\rm max}$ is $N_{\rm max} = 22$.
The typical value of the Gaussian width parameters
are $b_{0} = 0.1$ fm and $\gamma = 1.3$.
Hence, the maximum  size of the width parameter becomes
$b = b_{0} \gamma^{N_{\rm max}-1} = 0.1 \times 1.3^{21} \simeq 25$ fm.

In GSM, the continuum is discretized with 40 points
for each partial wave and the maximum momentum 
is $k_{\rm max} = 3.5$ fm$^{-1}$.

\subsection{$^{6}$He in the  $^{4}$He+$2N$ model space}

First, we show results for the ground state $0^{+}_{1}$
and the first excited state $2^{+}_1$ of $^{6}$He.
The ground state of $^{6}$He is bound one with an energy $E=0.97$ MeV
from the $^{4}$He+$2n$ threshold.
Hence, we can take the rotation angle as $\theta = 0$ for the calculation
of this state in GEM+CS approach.

\begin{table}
  \caption{Energies of the ground $0_1^{+}$ and
    the first excited $2_1^{+}$ states of $^{6}$He
    calculated using the GEM+CS and  GSM approaches.
   All units except for the angular momentum are in MeV.}
  \center
  \begin{tabular}{cccccc}
   \hline
    \hline
   &  $\ell_{\rm max} $  & GEM+CS & & GSM \\
    \hline
    & 1 & $-0.117 $  & & $-0.116   $   \\
    & 2 & $-0.737  $  & & $-0.737  $  \\
  $E(0^{+}_{1})$  & 3 & $-0.870  $ &  & $-0.870  $ \\
    & 4 & $-0.933  $  & & $-0.932  $  \\
    & 5 & $-0.978  $  & & $-0.977  $  \\
    \hline
    & $\ell_{\rm max} $  & GEM+CS & & GSM \\
        \hline
    & 1 & $0.805  - i 0.086 $ & & $0.804 - i 0.086$  \\
    & 2 & $0.675  - i 0.038 $ & & $0.669 - i 0.041$  \\
   $E(2^{+}_{1})$ &  3 & $0.628  - i 0.027 $ & & $0.619 - i 0.030$  \\
    & 4 & $0.605  - i 0.023 $ & & $0.595 - i 0.026$  \\
    & 5 & $0.589 -  i 0.021 $ & & $0.577 - i 0.024$   \\
    \hline
    \hline
  \end{tabular}
   \label{tab:6He_energies_0+_2+}
 \end{table}

We calculate energies of $^{6}$He by changing the maximum angular momentum for the coordinates
$\bv{r}_{1}$ and $\bv{r}_{2}$ from $\ell_{\rm max} =1$ to $5$,
where $\ell_{\rm max}$ is the maximum 
angular momentum in the basis function for the $^{4}$He+$n$
subsystem.
Parameters of the interaction are chosen to reproduce
the binding energy of the ground state of $^{6}$He in a model 
space with $\ell_{\rm max} = 5$.

 \begin{figure}[htpb]
  \includegraphics{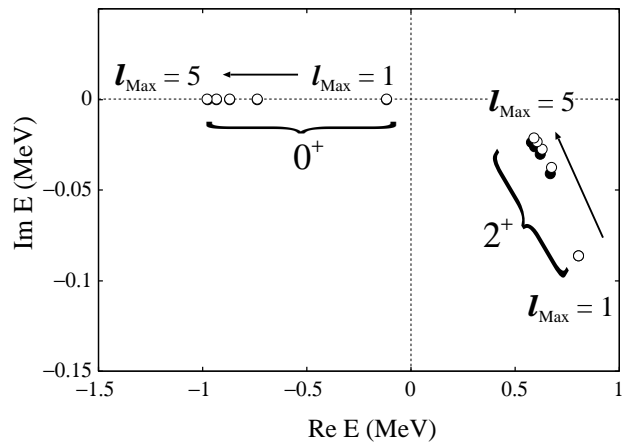}
  \caption{Convergence of the poles of the ground $0_1^{+}$ and the
    first excited $2_1^{+}$ states of $^{6}$He, which are 
      calculated using the GEM+CS approach and the GSM for $1\leq \ell_{\rm max}\leq 5$.
      Open and solid circles denote GEM+CS and GSM results, respectively.}
    \label{fig:Pole_He}
\end{figure}

The energies of the $0^{+}_{1}$ state are shown in
Table~\ref{tab:6He_energies_0+_2+} for different 
values of $\ell_{\rm max}$.
One can see that the calculation for $\ell_{\rm max}=1$, which includes 
the $s_{1/2}$-, $p_{3/2}$- and $p_{1/2}$-orbits of the $^{4}$He+$N$ system,
is not enough to reproduce the binding energy of $^{6}$He.
The inclusion of higher angular momenta ($\ell_{\rm max} \geq 2$) 
improves the calculated energy significantly.
Nevertheless, even $\ell_{\rm max}=5$ is not enough to 
obtain the converged ground state energy since the T-type
Jaccobi configuration of valence neutrons 
is very important~\cite{Ao95_PTP93}.
However, since the scope of this paper is to compare results of 
GEM+CS approach and GSM, we restrict the maximum angular momentum for the core+$N$ system 
to $\ell_{\rm max} = 5$ and determine the interaction parameters in this model space.

We find a good agreement between GEM+CS and GSM for a Borromean
$^{6}$He nucleus.
The $\ell_{\rm max}$-dependence of the $0^+_1$ and $2^+_1$ energies
is shown in Table~\ref{tab:6He_energies_0+_2+}  and Fig.~\ref{fig:Pole_He}.

\begin{figure}[htpb]
   \includegraphics{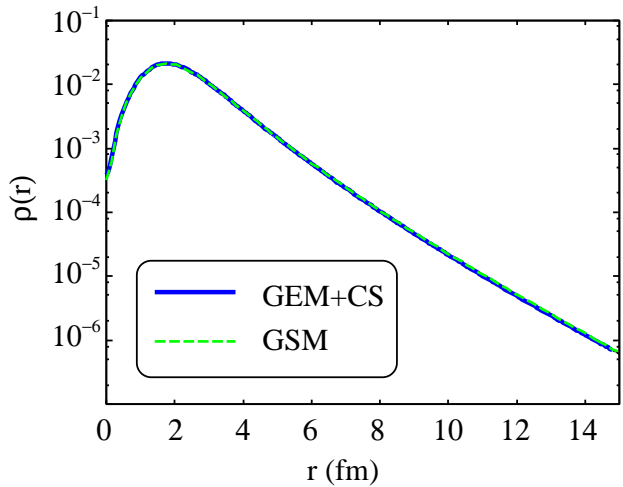}
   \caption{(Color online)
     The density of valence neutrons in COSM coordinate system
      for the ground $0_1^{+}$ state of $^{6}$He (color online).
      The normalization of the density distribution is 1. }
    \label{fig:Density_He}
\end{figure}

The density of valence neutrons in the $0_1^+$ state of $^6$He
is plotted in Fig.~\ref{fig:Density_He}. 
One can see that the GEM+CS and GSM approaches give indistinguishable
results for the density distributions in the $0_1^+$ halo
configuration of $^{6}$He.

Results for the $2^{+}_{1}$ narrow resonance are shown
in Table~\ref{tab:6He_energies_0+_2+}  and Fig.~\ref{fig:Pole_He}.
The difference between GEM+CS and GSM results in this case is at most $\sim$10 keV.
The trajectories of the $2^{+}_{1}$ state of the GEM+CS and GSM poles
are shown in Fig.~\ref{fig:Pole_He}. 
Similarly, as for the $0^{+}_{1}$-state, results
of the GEM+CS and GSM approaches agree well.

\subsection{$^{6}$Be in the $^{4}$He+$2N$ model space}

The $^{6}$Be nucleus, the mirror system of $^{6}$He, is unbound in the ground state.
In this section, we shall 
compare results of GEM+CS and GSM for the $0_1^+$ and $2_1^+$
states of $^{6}$Be described as a $^{4}$He+$2p$ three-body system. 

\begin{figure}[htpb]
  \includegraphics{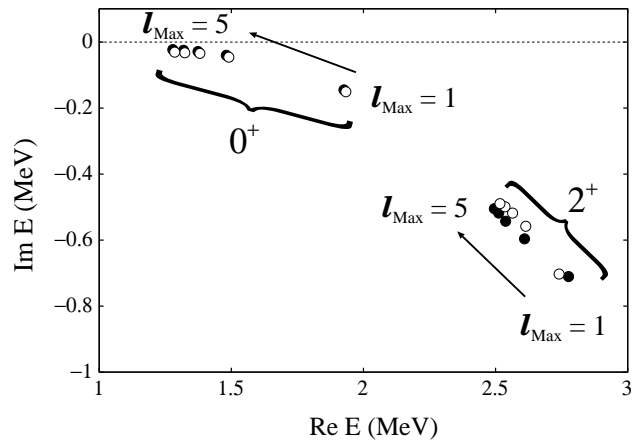}
  \caption{Poles of the ground and first excited states
    of $^{6}$Be calculated using GEM+CS approach 
    and GSM from $\ell_{\rm max}=1$ to $5$.
      Open and solid circles correspond to GEM+CS and GSM results, respectively.}
    \label{fig:Pole_Be}
\end{figure}

Calculated energies of the $0^{+}_{1}$ and $2^{+}_{1}$ states
for different $\ell_{\rm max}$ values 
are shown in Table \ref{tab:6Be_energies_0+_2+}.
The difference of GEM+CS and GSM energies  is less than $\sim$10 keV
for the $0^{+}_{1}$ state and $\sim$20 keV for the $2^{+}_{1}$ state.

\begin{table} 
  \caption{Energies of the ground $0^{+}_{1}$ and the first excited
    $2^{+}_{1}$ states of $^{6}$Be calculated 
    using the GEM+CS and GSM approaches.
   All units except for the angular momentum are in MeV.}
 \center
  \begin{tabular}{cccccc}
   \hline
   \hline
    &
   $\ell_{\rm max} $  & GEM+CS && GSM \\
    \hline
    & 1 & $1.932 -i0.152$  &&  $1.926 - i 0.146$ \\
    & 2 & $1.490 -i0.046$  &&  $1.482 - i 0.041$ \\
  $E(0^{+}_{1})$  & 3 & $1.380 -i0.036$ & &  $1.374 - i 0.030$ \\
    & 4 & $1.324 -i0.033$  &&  $1.318 - i 0.026$ \\
    & 5 & $1.285 -i0.031$  &&  $1.279 - i 0.024$ \\
    \hline
     &
    $ \ell_{\rm max} $  & GEM+CS && GSM \\
    \hline
    & 1 & $2.741 -i0.703  $ && $2.776 - i 0.711 $ \\
    & 2 & $2.614 -i0.559  $ && $2.610 - i 0.596 $ \\
   $E(2^{+}_{1})$ & 3 & $2.565 -i0.518  $ && $2.538 - i 0.543 $ \\
    & 4 & $2.537 -i0.500  $ && $2.512 - i 0.518 $ \\
    & 5 & $2.517 -i0.491  $ && $2.495 - i 0.505 $ \\  
    \hline
    \hline
  \end{tabular}
   \label{tab:6Be_energies_0+_2+}
\end{table}

The trajectory of the $0_1^+$ and $2_1^+$ poles in the complex energy
plane is shown in Fig.~\ref{fig:Pole_Be}.
Contrary to the $0^{+}_{1}$ state, one may notice a slight difference between trajectories of 
$2^{+}_{1}$-poles in GEM+CS approach and in GSM.
This difference diminishes with increasing $\ell_{\rm max}$.

\subsection{Discussion}

In the comparison between the GEM+CS and GSM approaches,
we obtain a good agreement for the bound state $0^{+}_{1}$ in $^{6}$He,
and narrow resonances; $2^{+}_{1}$ in $^{6}$He and $0^{+}_{1}$ in $^{6}$Be. 
A small difference appears only for the $2^{+}_{1}$ broad resonance in
$^{6}$Be. Below, we shall discuss a possible origin of
such a small difference in the numerical results.

Both GEM+CS and GSM approaches solve the non-Hermitian problem. 
In the GEM+CS approach, the wave function of a resonance becomes
${\cal L}^{2}$-integrable  with the help of the complex rotation.
As a result, the Hamiltonian becomes non-Hermitian.
The standard procedure to find the optimum values of the parameters
is to search for a stationary point of the eigenvalue
with respect to the variational parameters.
The variational parameters are 
the complex rotation angle $\theta$ and  the parameter $b_{0}$
in a definition of the Gaussian width;
$b_{n_{i}} = b_{0} \gamma^{n_{i}-1}$~\cite{Ao06_PTP116}
for the Gaussian basis functions.
The optimization procedure is a simplified version of the generalized variational
principle for complex eigenvalues~\cite{Moiseyev}.
The procedure works efficiently and gives very accurate solutions even for 
broad resonant states~\cite{Ao06_PTP116}.

\begin{figure}[htpb]
  \includegraphics{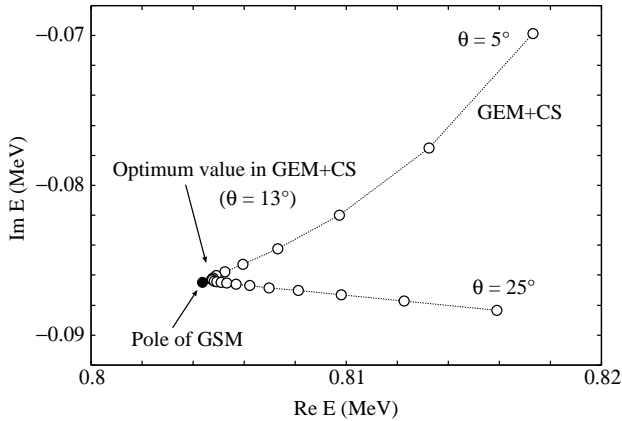}
  \caption{Poles of $^{6}$He $(2^{+})$ with $\ell_{\rm max} = 1$.
    For GEM+CS, we change the rotation angle $\theta$
    from $5^{\circ}$ to $25^{\circ}$ in step of $1^{\circ}$. }
    \label{fig:Pole_COSM_GSM_6He_2n}
\end{figure}

GSM is formulated in the Berggren set,
which includes bound single-particle states, single-particle resonances
and scattering states from the discretized contour for each considered $(\ell,j)$. 
Consequently, the Hamiltonian matrix in this basis becomes
complex-symmetric.
The number of scattering states on each discretized contour 
$(\ell,j)$ and the momentum cutoff have to be chosen to assure
the completeness of many-body calculations.
Moreover, in the HO expansion procedure of calculating the TBMEs,
the dependence on the oscillator length and the number
of oscillator shells should be carefully examined.

Fig.~\ref{fig:Pole_COSM_GSM_6He_2n} presents
a trajectory of the $2^{+}_{1}$ narrow resonant pole of  $^{6}$He calculated
in the GEM+CS approach by changing the rotation angle $\theta$,
where the stationary point at the optimum value of the rotation 
angle is $\theta_{\rm opt} = 13^{\circ}$,
and the optimum point for the GSM calculation,
which is obtained with the oscillator length and the number of
oscillator shells are $b_{\rm HO} = 2$ fm and $N = 41$, respectively.
In this case, the difference is only $\sim 1$ keV,
and both methods give almost the equivalent result.

On the other hand, the $2^{+}_{1}$ state of $^{6}$Be is a broad
resonant pole due to the presence of the Coulomb force
for all three particles.
The optimum value of the rotation angle in GEM+CS calculation is
$\theta_{\rm opt} = 17^{\circ}$, and the optimal HO oscillator length
in GSM calculations is $b_{\rm HO} = 3$ fm.
The difference of complex GEM+CS and GSM eigenenergies
becomes in the order of 10 keV.
To improve the agreement for the eigenvalues obtained
by GEM+CS and GSM,
it would be necessary to examine the optimization of
the variational parameters more precisely.
However, in the practical point of view,
the difference is only less than 1 percent to the total energy.

The convergence can be tested by introducing an extrapolation
procedure, e.g. the Richardson extrapolation~\cite{Ri1911_PTRSL210}.
We extrapolate the energy $E$ as a function of $1/\ell_{\rm max}$
to $1/\ell_{\rm max} = 0$.
The energies $E(1/\ell_{\rm max})$ of the $2^{+}$-state of $^{6}$Be
become $2.483-i0.474$  and $2.464 - i0.481$ (MeV)
for GSM+CS and GSM, respectively.
The difference becomes smaller than that of the $\ell_{\rm max} = 5$ case.
Hence, we can conclude that both methods provide a sufficient
accuracy even for the calculation of the broad resonant states.

\section{Summary}
GSM and GEM+CS are two different theoretical approaches
which allow to describe unbound 
resonant states. These two approaches differ in the choice of the basis functions and the 
numerical procedure to obtain the eigenvalues. To benchmark GSM and GEM+CS approaches, 
we have performed a precise comparison for weakly bound and unbound states using the same 
Hamiltonian in the COSM coordinates preserving the translational invariance.
For a weakly bound ground-state of $^{6}$He,
GSM and GEM+CS give essentially identical results. For the three-body 
resonance states, GEM+CS and GSM give very close results
proving the reliability of both schemes 
of the calculation for unbound states.
The slight difference between GSM and GEM+CS results for broad resonances may have different 
origins. The HO expansion procedure in calculating the TBMEs
in GSM may lead to rounding errors,  especially for broad many-body resonances.
On the other hand, the stationarity condition in GEM+CS 
approach could also be a source of small imprecision for broad
resonances.
Based on our results, 
we conclude that both approaches are essentially equivalent for all quantities studied. 
The other work for a comparison in the $^{6}$He system has been done
and also shows a good agreement between two different approaches~\cite{CS_preprint}.

\begin{acknowledgements}
  We would like to thank W. Nazarewicz and
  members of the nuclear theory group at Hokkaido University
  for fruitful discussions.
  This work was supported by the Grant-in-Aid for
  Scientific Research (No. 21740154) from the Japan Society
  for the Promotion of Science, and FUSTIPEN
  (French-U.S. Theory Institute for Physics with Exotic Nuclei)
  under DOE grant number DE-FG02-10ER41700.
\end{acknowledgements}

\end{document}